\documentclass[12pt]{article}
\usepackage{epsf}
\usepackage{amsmath,amssymb}

\usepackage[dvips]{graphicx}

\usepackage{comment}

\begin{document}

\newcommand{\lsim}{\stackrel{<}{_\sim}}
\newcommand{\gsim}{\stackrel{>}{_\sim}}

\newcommand{\rem}[1]{{$\spadesuit$\bf #1$\spadesuit$}}

\renewcommand{\thefootnote}{\fnsymbol{footnote}}
\setcounter{footnote}{0}

\begin{titlepage}

\def\thefootnote{\fnsymbol{footnote}}

\begin{center}

\hfill UT-13-41\\
\hfill December, 2013\\

\vskip .75in

{\large \bf 

  Cosmic-Ray Neutrinos from the Decay of Long-Lived Particle\\
  and the Recent IceCube Result

}

\vskip .75in

{\large 
Yohei Ema, Ryusuke Jinno and Takeo Moroi
}

\vskip 0.25in

\vskip 0.25in

{\em Department of Physics, University of Tokyo,
Tokyo 113-0033, Japan}

\end{center}
\vskip .5in

\begin{abstract}

  Motivated by the recent IceCube result, we study high energy
  cosmic-ray neutrino flux from the decay of a long-lived particle.
  Because neutrinos are so transparent, high energy neutrinos produced
  in the past may also contribute to the present neutrino flux.  We
  point out that the PeV neutrino events observed by IceCube may
  originate in the decay of a particle much heavier than PeV if its
  lifetime is shorter than the present cosmic time.  It is shown that
  the mass of the particle responsible for the IceCube event can be as
  large as $\sim 10^{10}\ {\rm GeV}$.  We also discuss several
  possibilities to acquire information about the lifetime of the
  long-lived particle.

\end{abstract}

\end{titlepage}

\renewcommand{\thepage}{\arabic{page}}
\setcounter{page}{1}
\renewcommand{\thefootnote}{\#\arabic{footnote}}
\setcounter{footnote}{0}

Neutrino astronomy provides a new window to the early universe.  This
is particularly because, contrary to particles with electromagnetic
interactions (i.e., $e^\pm$, $p$ and $\bar{p}$, $\gamma$ and so on),
neutrinos have very weak interaction and are very transparent.  Thus,
the cosmic-ray neutrino spectrum in the present universe contains
various information about the production processes of energetic
neutrinos.

Recently, the IceCube experiment reported the result of their analysis
on high-energy neutrino events \cite{Aartsen:2013pza}. The IceCube
experiment observed $28$ high-energy neutrino events with $E_{\rm
  EM}>30\ {\rm TeV}$ (with $E_{\rm EM}$ being the deposited
electromagnetic-equivalent energy in detector), which is substantially
larger than the number of events expected from the atmospheric
backgrounds (which is $10.6^{+5.0}_{-3.6}$ events).  Notably, the
IceCube experiment found two events with deposit energy of $\sim {\rm
  PeV}$ ($1041^{+132}_{-144}\ {\rm TeV}$ and $1141^{+143}_{-133}\ {\rm
  TeV}$, nicknamed as Ernie and Bert, respectively), while no event
with larger $E_{\rm EM}$ has been observed.

IceCube claims the existence of a source of high energy cosmic-ray
neutrinos other than the atmospheric one \cite{Aartsen:2013pza}.  We
may consider particle-physics or astrophysical origin of such high
energy neutrinos, the former of which is the subject of our study.
(Possible astrophysical origins include Active Galactic Nuclei
\cite{Essey:2009ju, Kalashev:2013vba, Stecker:2013fxa}, $\gamma$-ray
burst \cite{Cholis:2012kq, Murase:2013ffa, Razzaque:2013dsa,
  Winter:2013cla}, hypernova remnants \cite{Fox:2013oza, Liu:2013wia},
star-forming galaxies \cite{Murase:2013rfa}, 
Galactic cosmic-rays \cite{Gupta:2013xfa, Gonzalez-Garcia:2013iha, Ahlers:2013xia},
neutron-star mergers \cite{Gao:2013rxa} and cosmogenic neutrinos
\cite{Roulet:2012rv, Laha:2013lka}.  For other astrophysical
discussion, see also \cite{Anchordoqui:2013qsi, He:2013zpa}.)
Although it is premature to make any conclusion, the negative
observation of the events with larger energy deposit may indicate that
the cosmic-ray neutrino spectrum has a cutoff at the energy around
$\sim {\rm PeV}$.  In fact, we also note that no event is observed in
the energy bins between $0.4-0.63\ {\rm PeV}$ and $0.63-1\ {\rm PeV}$.
It may be a consequence of a peak of the cosmic-ray electron neutrino
spectrum at $\sim {\rm PeV}$; this is because, within experimental
uncertainties, the deposited energy is equal to the energy of the
initial-state neutrino for $\nu_e$ charged current events, while it is
below the energy of the neutrino for other types of events.

From particle-physics point of view, the structure in the neutrino
spectrum mentioned above may be realized with a new physics at the
energy scale higher than $\sim {\rm PeV}$.  We consider such a case in
this letter.  In particular, we study whether the decay of a new
particle $X$ with its mass $m_X\gtrsim 1\ {\rm PeV}$ is responsible
for the high-energy neutrino events observed by IceCube.  For example,
if neutrinos are produced by the decay of a particle $X$ with $m_X\sim
{\rm PeV}$ in the present universe, a neutrino spectrum with the
cutoff at $\sim {\rm PeV}$ may be obtained.  The case where the
dark-matter particle plays the role of $X$ was considered in
\cite{Feldstein:2013kka, Esmaili:2013gha, arXiv:1311.5864}.  (For
early discussion about related issues, see \cite{Datta:2004sr,
  Anchordoqui:2005is, Aloisio:2006kk, Allahverdi:2009se,
  Blennow:2009ag, Lindner:2010rr, Esmaili:2012us}.)  On the contrary,
even if the mass of $X$ is much larger than $\sim {\rm PeV}$, there
still exists a possibility to produce present cosmic-ray neutrinos
with $E\sim {\rm PeV}$.  This is because, if the decay of $X$ occurs
earlier than the present epoch, the energy of the neutrino produced by
the $X$ decay is redshifted.

In this letter, we study cosmic-ray neutrinos produced by a long-lived
particle $X$.  We show that some of the high energy neutrino events
observed by IceCube can be due to neutrinos produced by the decay of
$X$.  In particular, we discuss that the peak in the cosmic-ray
neutrino spectrum may show up at $E\sim 1\ {\rm PeV}$ even with the
mass of $X$ much larger than PeV, if the lifetime of $X$ is shorter
than the present age of the universe.  For such a scenario to work,
the decay of $X$ is required to occur at $z\lesssim O(10^3)$ (with $z$
being the redshift), which implies that the mass of $X$ can be as
large as $10^{10}\ {\rm GeV}$.  We also discuss how the neutrino
spectrum depends on the properties of $X$.

We start our discussion without assuming any particular model for the
heavy particle $X$.  Instead, we parametrize the properties of $X$
with the following three quantities:
\begin{align}
  m_X,~~~\tau_X,~~~Y_X,
\end{align}
where $m_X$ and $\tau_X$ are the mass and the lifetime of $X$,
respectively.  In addition, $Y_X$ is the so-called yield variable
\begin{align}
  Y_X \equiv \left[ \frac{n_X(t)}{s(t)} \right]_{t\ll \tau_X},
\end{align}
with $s$ being the entropy density; with $Y_X$, the number density
of $X$ is given by
\begin{align}
  n_X (t) = Y_X s(t) e^{-t/\tau_X}.
\end{align}
In the following, to make our point clearer, we concentrate on the
case where the neutrino produced by the decay of $X$ is monochromatic
(with the energy $\bar{E}_\nu$); the energy distribution of the
neutrino produced by the decay of $X$ is expressed as
\begin{align}
  \frac{d N_\nu^{(X)}}{d E} = 
  \bar{N}_\nu \delta(E-\bar{E}_\nu),
\end{align}
where $\bar{N}_\nu$ is the number of the neutrino produced by the
decay of one $X$.  (For our numerical study, we take
$\bar{E}_\nu=m_X/2$ and $\bar{N}_\nu=1$.)  Because we are interested
in electron neutrino, with which the peak in the IceCube result may be
explained, we assume that the decay of $X$ produces a sizable amount
of $\nu_e$ (after taking account of the effects of neutrino
oscillation).  With the monochromatic distribution, we will see that
the peak in the cosmic-ray neutrino flux can be obtained.

Now, we discuss the flux of the cosmic-ray neutrinos produced by the
decay of $X$.  In the total flux, there exist two contributions:
\begin{align}
  \Phi_\nu (t, E)
  = \Phi_\nu^{\rm (Cosmo)} (t, E) +
  \Phi_\nu^{\rm (Galaxy)} (t, E),
\end{align}
where $\Phi_\nu^{\rm (Cosmo)}$ and $\Phi_\nu^{\rm (Galaxy)}$ are
contributions from cosmological distance and from our Galaxy,
respectively.  (The neutrino number density is given by $n_\nu(t)=\int
dE \Phi_\nu (t,E)$.)

The flux of high energy neutrinos from the cosmological distance obeys
the following Boltzmann equation:
\begin{align}
  \frac{\partial \Phi_\nu^{\rm (Cosmo)}}{\partial t} = 
  - 2 H \Phi_\nu^{\rm (Cosmo)} 
  + H E \frac{\partial \Phi_\nu^{\rm (Cosmo)}}{\partial E}
  + S_\nu (t,E)
  - \gamma_\nu (t,E) \Phi_\nu^{\rm (Cosmo)},
  \label{Boltzmann}
\end{align}
where $H$ is the expansion rate of the universe, $S_\nu (t,E)$ is
the source term:
\begin{align}
  S_\nu (t,E) = \frac{1}{4\pi} 
  \frac{n_X (t)}{\tau_X}  \frac{d N_\nu^{(X)}}{d E},
\end{align}
with $n_X$ being the number density of $X$, and $\gamma_\nu (t,E)$ is
the scattering rate:
\begin{align}
  \gamma_\nu (t,E) =
  \frac{1}{16 \pi^2 E^{2}}
  \int_0^\infty d k
  f_{\rm BG} (k)
  \int_0^{4kE} ds s \sigma_{\rm tot}(s),
\end{align}
with $f_{\rm BG}$ being the distribution function of the background
(target) particle and $\sigma_{\rm tot}(s)$ the total scattering cross
section of high energy neutrinos for the processes with the
center-of-mass energy $\sqrt{s}$.  By solving the Boltzmann equation,
the neutrino spectrum at the present epoch $t_0$ is given by
\begin{align}
  \Phi_\nu^{\rm (Cosmo)} (t_0, E) = 
  \int_{-\infty}^{t_0} dt
  \left( \frac{a_0}{a(t)} \right)^{-2}
  D_\nu (E; z(t))
  S_\nu (t, a_0 E / a(t)),
\end{align}
where $a$ is the scale factor and $a_0 \equiv a(t_0)$.  In addition,
\begin{align}
  D_\nu (E; z(t)) = \exp 
  \left[ 
    -\int_t^{t_0} dt' 
    \gamma_\nu (t', (1+z(t'))E)
  \right],
  \label{D_nu}
\end{align}
with
\begin{align}
  1+z(t) \equiv \frac{a_0}{a(t)}.
\end{align}
(Notice that the arguments of $D_\nu (E;z)$ are chosen to be the
present neutrino energy $E$ and the redshift at the time of the
neutrino production $z$.)  When the decay process produces
monochromatic neutrino, we obtain
\begin{align}
  \Phi_\nu^{\rm (Cosmo)} (t_0, E) = 
  \frac{1}{4\pi} 
  \frac{\bar{N}_\nu Y_X s(t_0) }{\tau_X E}
  \left[ 
    \frac{ e^{-\bar{t}/\tau_X} D_\nu (E; z(\bar{t}))}
    { H (\bar{t}) }
  \right]_{1+z(\bar{t})=\bar{E}_\nu/E}.
  \label{Phi_nu}
\end{align}
In our analysis, the effects of secondary neutrinos produced by the
scattering processes are not included.  This can be justified as far
as the damping factor $D_\nu (E; z)$ is close to $1$.  Hereafter, we
mostly consider such a case.

In the case of our interest, important scattering processes of the
high energy neutrinos are with background neutrinos, and hence $f_{\rm
  BG}(k) = (e^{k/T_\nu}+1)^{-1}$, where $T_\nu$ is the neutrino
temperature and is related to the temperature of the background
radiation as $T_\nu=(4/11)^{1/3}T$.  
Here we neglect the neutrino masses in the distribution function, since, 
as we will see later, the damping factor becomes effective only for 
$z \gtrsim O(10^3)$, where $T_\nu \gtrsim O(0.1\ {\rm eV})$. 
Then, the scattering rate is
given by
\begin{align}
  \gamma_\nu (t,E) = 
  \frac{T_\nu(t)}{\pi^2} 
  \int_0^\infty d k
  k \log \left( 1 + e^{-k/T_\nu(t)} \right)
  \sigma_{\rm tot}(s = 4kE).
\end{align}
For the calculation of the damping
rate of $\nu_\ell$, $\sigma_{\rm tot}$ is obtained by taking account
of the effects of the following processes:
\begin{itemize}
\item $\nu_\ell + \bar{\nu}_{\ell, {\rm BG}}
  \rightarrow \nu_\ell + \bar{\nu}_\ell$,
\item $\nu_\ell + \bar{\nu}_{\ell, {\rm BG}}
  \rightarrow \ell + \bar{\ell}$,
\item $\nu_\ell + \bar{\nu}_{\ell, {\rm BG}}
  \rightarrow f + \bar{f}$, with
  $f\neq\nu_\ell$, $\ell$,
\item $\nu_\ell + \bar{\nu}_{\ell',{\rm BG}}
  \rightarrow \nu_\ell + \bar{\nu}_{\ell'}$,
  with $\ell\neq\ell'$,
\item $\nu_\ell + \bar{\nu}_{\ell',{\rm BG}}
  \rightarrow \ell + \bar{\ell}'$,
  with $\ell\neq\ell'$,
\item $\nu_\ell + \nu_{\ell, {\rm BG}} \rightarrow \nu_\ell + \nu_{\ell}$.
\item $\nu_\ell + \nu_{\ell',{\rm BG}}
  \rightarrow \nu_\ell + \nu_{\ell'}$,
  with $\ell\neq\ell'$,
\end{itemize}
where $f$ denotes quarks and leptons, $\ell$ denotes charged leptons,
and the subscript ``BG'' is for background neutrinos.  (Cross sections
for these processes are given in Appendix.)  As we will see, for a
cosmic-ray neutrino whose present energy is $E\sim 1\ {\rm PeV}$, the
damping becomes important only for $z\gtrsim O(10^{3})$, for which the
typical center-of-mass energy of the scattering process is
$\sqrt{s}\gtrsim 1\ {\rm GeV}$.  Therefore, in our calculation of
$\sigma_{\rm tot}$, we neglect the masses of leptons as well as those
of first- and second-generation quarks.  We have checked that the
damping factor $D_\nu (E;z)$ is insensitive to the behavior of
$\sigma_{\rm tot}(s)$ with $\sqrt{s}\lesssim$ a few GeV.

\begin{figure}[t]
  \centerline{\epsfxsize=0.55\textwidth\epsfbox{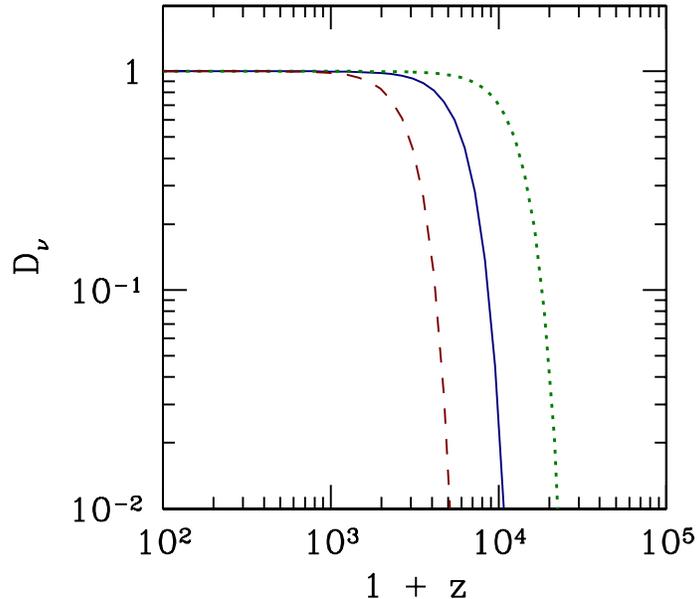}}
  \caption{\small The damping factor $D_\nu (E;z)$ defined in Eq.\
    \eqref{D_nu} as a function of the redshift $1+z$.  The present
    energy of the neutrino is $E=0.1$ (green-dotted), $1$ (blue-solid),
    and $10\ {\rm PeV}$ (red-dashed) from right to left.}
  \label{fig:optdepth}
\end{figure}

In Fig.\ \ref{fig:optdepth}, we plot the damping factor defined in
Eq.\ \eqref{D_nu} as a function of the redshift.  
For the present neutrino energy of $E\sim 1\
{\rm PeV}$, $D_\nu(E;z)$ is significantly suppressed for $z\gtrsim
O(10^3)$.  For example, $D_\nu (E=1\ {\rm PeV};z)<0.5$ for $z\gtrsim
6\times 10^3$.

\begin{figure}
  \centerline{\epsfxsize=0.95\textwidth\epsfbox{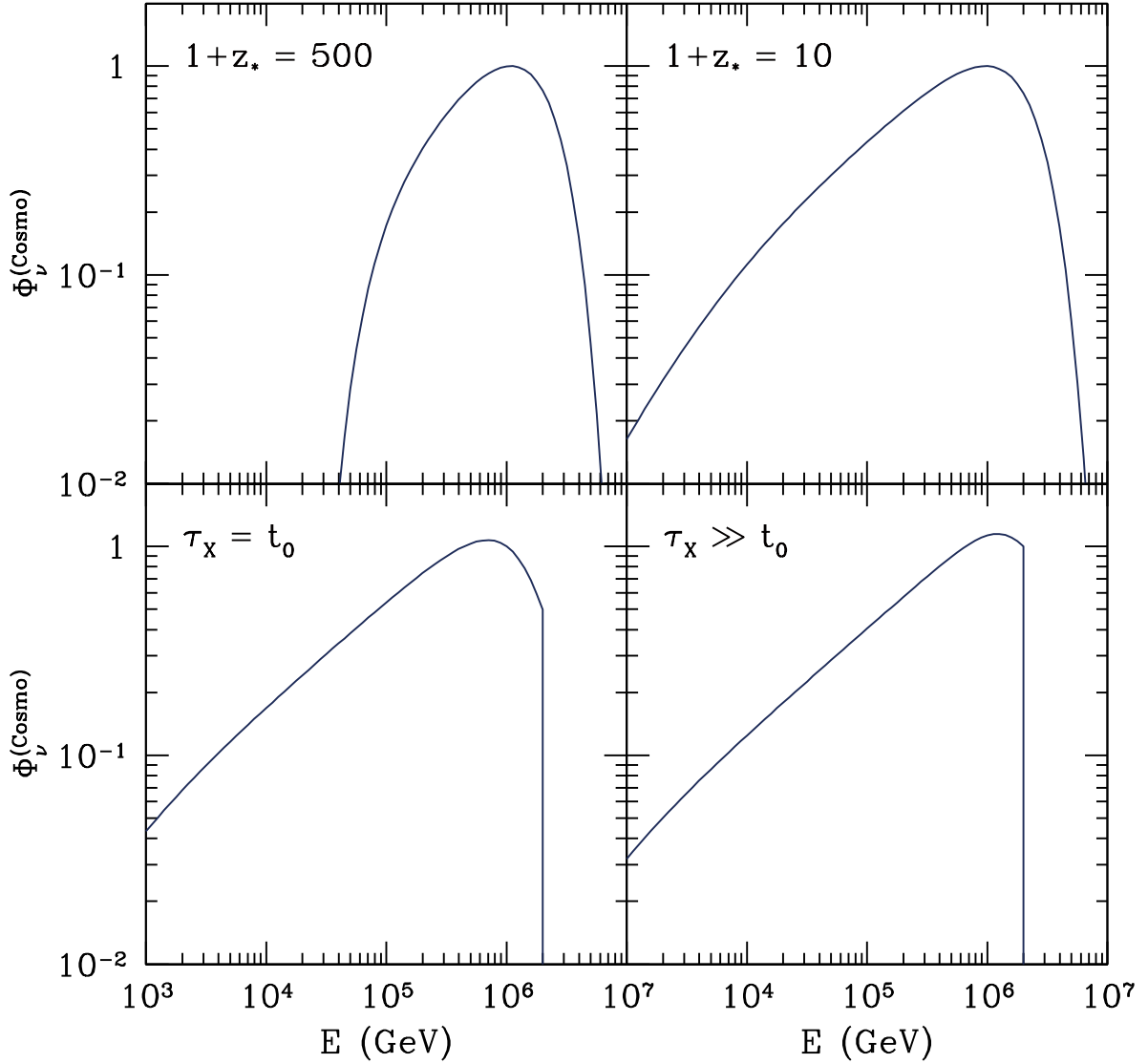}}
  \caption{\small Present cosmic-ray neutrino flux $\Phi_\nu^{\rm
      (Cosmo)}$ given in Eq.\ \eqref{Phi_nu} as a function of the
    present energy $E$. Here, we take $(\tau_X,\bar{E}_\nu)
    =(4.2\times 10^{13}\ {\rm sec},1000\ {\rm PeV})$ (top-left, which
    corresponds to $1+z_*=500$), $(1.7\times 10^{16}\ {\rm sec},20\
    {\rm PeV})$ (top-right, which corresponds to $1+z_*=10$), $(t_0,2\
    {\rm PeV})$ (bottom-left), and $(10^{29}\ {\rm sec},2\ {\rm PeV})$
    (bottom-right). The normalization is arbitrary.}
    
  \label{fig:Phinu_zdep}
\end{figure}

In Fig.\ \ref{fig:Phinu_zdep}, we plot $\Phi_\nu^{\rm (Cosmo)}(E)$
given in Eq.\ \eqref{Phi_nu} for several values of $\tau_X$.
(Hereafter, we concentrate on the flux at present, so we omit $t_0$
from the argument.)  The normalization of the spectrum in Fig.\
\ref{fig:Phinu_zdep} is arbitrary.  One can easily see a peak in the
spectrum; the position of the peak depends on the initial energy of
the neutrino as well as on the lifetime of $X$.  If $\tau_X\ll t_0$
and $z_*\lesssim 10^3$ (for which the damping factor is negligible),
the position of the peak is approximately given by
\begin{align}
  E_{\rm peak}^{\rm (Cosmo)}
  \simeq 
  0.5 \times \frac{\bar{E}_\nu}{1 + z_*}: ~~~
  \mbox{for $\tau_X\ll t_0$},
\end{align}
where
\begin{align}
  z_* = z(\tau_X).
\end{align}
This behavior can easily be understood by using the fact that most of
$X$ decays at the cosmic time of $t\sim\tau_X$, and that the energy of
the emitted neutrino is redshifted by the factor of $\sim (1 +
z_*)^{-1}$.  One can also see that, compared to the flux for
$1+z_*=10$, that for $1+z_*=500$ shows a significant suppression for
$E\lesssim 0.1\ {\rm PeV}$.  This is due to the fact that the damping
factor $D_\nu$ becomes extremely small for $z\gtrsim O(10^3)$.  (The
flux for $1+z_*=10$ is almost unaffected by the damping.)  For
$1+z_*=500$, $D_\nu\simeq 0.5$ for $E= 0.08\ {\rm
  PeV}$.  For smaller $E$, secondary neutrinos produced by the
scattering processes may also contribute and the flux may be affected.
However, for such a redshift, the damping factor becomes close to $1$
for $E\gtrsim 0.1\ {\rm PeV}$, so we believe that Eq.\ \eqref{Phi_nu}
well describes the contribution from cosmological distance in the
energy region of our interest.  Contrary to the case of $\tau_X\ll
t_0$, the shape of the spectrum for $\tau_X\gg t_0$ is insensitive to
$\tau_X$ as far as $\bar{E}_\nu$ is fixed.  We can also see that the
shape of the spectrum for $\tau_X\gg t_0$ is quite different from that
for $\tau_X\ll t_0$.  In particular, $\Phi_\nu^{\rm (Cosmo)}$ has a
sharp edge at $E=\bar{E}_\nu$.

Next, we consider the contribution from the Milky-Way Galaxy.  Because
the Galactic contribution is not isotropic, we define
\begin{align}
  \tilde{\Phi}_\nu^{\rm (Galaxy)} (E,\hat{l}) = 
  \frac{1}{4\pi} \frac{\bar{N}_\nu}{\tau_X}
  \delta(E-\bar{E}_\nu)
  \int_{\rm l.o.s.} d \vec{l} 
  n_X (\vec{l}),
\end{align}
where l.o.s.\ stands for the line-of-sight integral, and $\hat{l}$
denotes the direction of the line-of-sight.  We assume that the
density of $X$ is proportional to that of dark matter:
\begin{align}
  n_X (\vec{l}) = \frac{1}{m_X} \frac{\Omega_X}{\Omega_{\rm DM}}
  \rho_{\rm DM} (\vec{l}),
\end{align}
where $\Omega_X$ and $\Omega_{\rm DM}$ are the density parameters of
$X$ and dark matter, respectively.  In addition, $\rho_{\rm DM}
(\vec{l})$ is the energy density of dark matter in the Galaxy; in our
numerical calculation, we adopt the NFW density profile
\cite{Navarro:1995iw, Navarro:1996gj}:
\begin{align}
  \rho_{\rm DM} (r) = \rho_\odot 
  \frac{r_\odot (r_{\rm c} + r_\odot)^2}{r(r_{\rm c} + r)^2},
\end{align}
where $r$ is the distance from the galactic center, $\rho_\odot\simeq
0.4\ {\rm GeV/cm^3}$ is the local halo density, $r_{\rm c} \simeq 20\
{\rm kpc}$ is the core radius, and $r_\odot \simeq 8.5\ {\rm kpc}$ is
the distance between the galactic center and the solar system. Then,
we define $\Phi_\nu^{\rm (Galaxy)}$ as the directional average of
$\tilde{\Phi}_\nu^{\rm (Galaxy)}$:
\begin{align}
  \Phi_\nu^{\rm (Galaxy)} (E) \equiv
  \frac{1}{4\pi}
  \int d \Omega_{\hat{l}}
  \tilde{\Phi}_\nu^{\rm (Galaxy)} (E,\hat{l}).
\end{align}

Now we are at the position to compare the predicted neutrino spectrum
with the IceCube result.  In our analysis, we particularly pay
attention to the fact that IceCube has observed two events in the bin
of $1 < E_{\rm EM}< 1.6\ {\rm PeV}$ while no event in $0.4 < E_{\rm
  EM}< 0.63\ {\rm PeV}$, $0.63 < E_{\rm EM}< 1\ {\rm PeV}$, and
$E_{\rm EM}>1.6\ {\rm PeV}$.  Thus, we concentrate on the case where
the spectrum of the neutrinos from $X$ decay, which is assumed to
contain a sizable fraction of $\nu_e$, becomes largest at $\sim {\rm
  PeV}$.  To make a comparison with the IceCube result, we define the
``averaged'' neutrino spectrum for $1 < E< 1.6\ {\rm PeV}$:
\begin{align}
  \bar{\Phi}_{\nu, {\rm PeV}} \equiv \frac{1}{0.6\ {\rm PeV}}
  \int_{1\ {\rm PeV}}^{\rm 1.6\ {\rm PeV}} dE
  \Phi_\nu (E),
  \label{PhiBar}
\end{align}
and similar quantities $\bar{\Phi}_{\nu, {\rm PeV}}^{\rm (Cosmo)}$ and
$\bar{\Phi}_{\nu, {\rm PeV}}^{\rm (Galaxy)}$ from $\Phi_\nu^{\rm
  (Cosmo)}$ and $\Phi_\nu^{\rm (Galaxy)}$, respectively.  For
$\tau_X\ll t_0$, $\bar{\Phi}_{\nu, {\rm PeV}}^{\rm (Galaxy)}$ is
negligibly small.  On the contrary, for $\tau_X\gg t_0$,
$\bar{\Phi}_{\nu, {\rm PeV}}^{\rm (Galaxy)}$ becomes more important
than $\bar{\Phi}_{\nu, {\rm PeV}}^{\rm (Cosmo)}$ (as far as $1<
\bar{E}_\nu<1.6\ {\rm PeV}$).  For $\bar{E}_\nu=1.1-1.6\ {\rm PeV}$,
$\bar{\Phi}_{\nu, {\rm PeV}}^{\rm (Cosmo)}/\bar{\Phi}_{\nu, {\rm
    PeV}}^{\rm (Galaxy)}\simeq 0.07-0.3$ when $\tau_X\gg t_0$.  Using
the fact that IceCube observed 2 events in the bin of $1 < E_{\rm EM}<
1.6\ {\rm PeV}$ within the live time of $662\ {\rm days}$, we estimate
\begin{align}
  \bar{\Phi}_{\nu, {\rm PeV}}^{\rm (IceCube)} \simeq
  3\times 10^{-16}
  \ {\rm m^{-2}\ sec^{-1}\ str^{-1}\ GeV^{-1}},
  \label{PhibarIceCube}
\end{align}
where we assumed that the total energy deposit is (almost) equal to
the energy of the initial-state neutrino, which is the case of the
charged current events of $\nu_e$.  Here, we used the effective area
of $15\ {\rm m^2}$ \cite{Aartsen:2013pza} as a reference value,
although this value also includes the effects of neutral current.  We
take this flux as a canonical value for our study.  We also note here
that, with the present setup, high-energy cosmic-ray neutrino events
with $E_{\rm EM}<1\ {\rm PeV}$ may also be induced by charged current
interactions of neutrinos other than $\nu_e$ and by neutral current
ones.  For the excess of the events with $E_{\rm EM}<1\ {\rm PeV}$, we
may also consider non-monochromatic initial neutrino injection as
another possibility \cite{arXiv:1311.5864}.

As we have mentioned, the neutrino flux is proportional to the yield
variable $Y_X$.  We estimate $Y_X$ which gives $\bar{\Phi}_{\nu, {\rm
    PeV}}=3\times 10^{-16} \ {\rm m^{-2}\ sec^{-1}\ str^{-1}\
  GeV^{-1}}$.  For $\tau_X\ll t_0$, for which $\bar{\Phi}_{\nu, {\rm
    PeV}}^{\rm (Galaxy)}$ is negligible, we choose $m_X$ so that
$E_{\rm peak}^{\rm (Cosmo)}=1.1\ {\rm PeV}$ (which is close to the
deposited energies of the most energetic events, i.e., Ernie and
Bert).  Then, the best-fit value of $Y_X$ to realize Eq.\
\eqref{PhibarIceCube} is given by
\begin{align}
  Y_X \simeq 1\times 10^{-26} \times \bar{N}_\nu^{-1} ~~~:~~~\tau_X\ll t_0.
  \label{Y_X(short)}
\end{align}
Notice that, for $\tau_X\ll t_0$, the best-fit value is insensitive to
the lifetime of $X$.  On the contrary, for $\tau_X\gg t_0$, we
choose $\bar{E}_\nu=1.1\ {\rm PeV}$ and obtain
\begin{align}
  Y_X \simeq 4\times 10^{-16} \times \bar{N}_\nu^{-1}
  \left( \frac{\tau_X}{10^{29}\ {\rm sec}} \right)
  ~~~:~~~\tau_X\gg t_0.
  \label{Y_X(long)}
\end{align}
In the case of $\tau_X\gg t_0$, $Y_X$ corresponds to the present yield
value of $X$.  Then, the present mass density of $X$ is estimated as
$\Omega_X\simeq 6\times 10^{14} \times Y_X (m_X/1\ {\rm PeV})$.
Combining this relation with Eq.\ \eqref{Y_X(long)}, we can see that
dark matter may play the role of $X$ if $\tau_X\sim O(10^{29}\ {\rm
  sec})$ \cite{Feldstein:2013kka, Esmaili:2013gha, arXiv:1311.5864}.
On the contrary, scenarios with longer lifetime do not work because of
the over-closure of the universe.

\begin{figure}[t]
  \centerline{\epsfxsize=0.55\textwidth\epsfbox{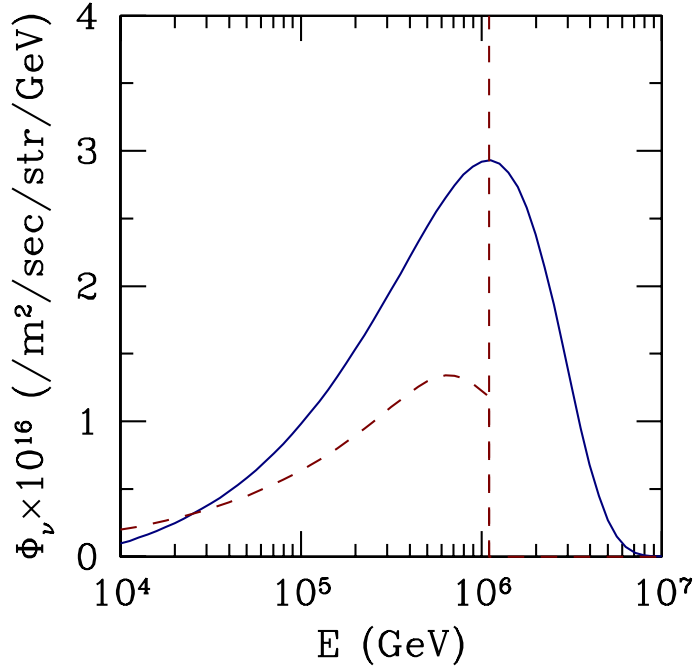}}
  \caption{\small The total present cosmic-ray neutrino flux from the
    long-lived particle $X$, normalized by $10^{-16}\ {\rm
      m^{-2}sec^{-1}str^{-1}GeV^{-1}}$. Here, we take
    $(\tau_X,\bar{E}_\nu,Y_X) =(5.2\times 10^{14}\ {\rm sec},2.2\times
    10^2\ {\rm PeV},1.2\times 10^{-26})$ (blue-solid, corresponding to
    $1+z_*=100$), and $(10^{29}\ {\rm sec},1.1\ {\rm PeV},4.0\times
    10^{-16})$ (red-dashed). The vertical line at $E= 1.1\ {\rm PeV}$
    is the contribution from the Galaxy.}
  \label{fig:PhinuIceCube}
\end{figure}

In Fig.\ \ref{fig:PhinuIceCube}, we show the neutrino spectrum for
$(\tau_X,\bar{E}_\nu)=(5.2\times 10^{14}\ {\rm sec},2.2\times 10^2\
{\rm PeV})$ and $(10^{29}\ {\rm sec},1.1\ {\rm PeV})$; the yield
variable is determined so that $\bar{\Phi}_{\nu, {\rm PeV}}$ is equal
to $\bar{\Phi}_{\nu, {\rm PeV}}^{\rm (IceCube)}$ given in Eq.\
\eqref{PhibarIceCube}.  We can see that the enhancement of the flux at
$E\sim 1\ {\rm PeV}$ is possible both for $\tau_X\ll t_0$ and
$\tau_X\gg t_0$.

One of the important check points of the present scenario, in
particular for the case of $\tau_X\ll t_0$, is the flux at the
``Glashow resonance'' \cite{Glashow:1960}.  For $E\simeq 6.3\ {\rm
  PeV}$, the event rate of IceCube for $\nu_e$ is enhanced because the
center-of-mass energy hits the $W$-boson pole
\cite{Anchordoqui:2004eb, Bhattacharya:2011qu}.  Indeed, the effective
area of the IceCube experiment for $E\simeq 6.3\ {\rm PeV}$ is about
40 times larger than that for $E\simeq 1\ {\rm PeV}$. One might wonder
if the present scenario is consistent with the negative observation of
the events for such an energy region because, if $\tau_X\ll t_0$, the
spectrum is non-vanishing even at $E\sim 6.3\ {\rm PeV}$.  Notably,
the flux for $E\gg E_{\rm peak}^{\rm (Cosmo)}$ is exponentially
suppressed because neutrinos with $E\gg E_{\rm peak}^{\rm (Cosmo)}$
are produced when $t\gg\tau_X$.  In fact, the flux at $E\simeq 6.3\
{\rm PeV}$ is quite sensitive to the value of $z_*$.  With the
position of the peak being fixed as $E_{\rm peak}^{\rm (Cosmo)}\simeq
1\ {\rm PeV}$, the ratio $\Phi_\nu^{\rm (Cosmo)} (6.3\ {\rm
  PeV})/\Phi_\nu^{\rm (Cosmo)} (E_{\rm peak}^{\rm (Cosmo)})$ becomes
smaller for larger value of $z_*$; this can be understood from the
fact that $\Phi_\nu^{\rm (Cosmo)} (E)$ is proportional to
$e^{-t_E/\tau_X}$, where $t_E$ is the time satisfying
$E=(1+z(t_E))^{-1}\bar{E}_\nu$ (see Eq.\ \eqref{Phi_nu}).  For the
case where $X$ decays in radiation- and matter-dominated epochs, for
example, this quantity is given by $e^{-t_E/\tau_X}=e^{-(E/E_*)^2}$
and $e^{-(E/E_*)^{3/2}}$, respectively, with $E_*\equiv
(1+z_*)^{-1}\bar{E}_\nu$.  For $E_{\rm peak}^{\rm (Cosmo)}=1\ {\rm
  PeV}$ ($1.1\ {\rm PeV}$), $\Phi_\nu^{\rm (Cosmo)} (6.3\ {\rm
  PeV})/\Phi_\nu^{\rm (Cosmo)} (E_{\rm peak}^{\rm (Cosmo)})$ is given
by $0.014$, $0.012$, and $0.001$, ($0.027$, $0.024$, and $0.003$) for
$1+z_*=10$, $100$, and $1000$, respectively.  Using the fact that
IceCube has not observed any event at around the Glashow resonance,
relatively large value of $z_*$ may be preferred.  However, the
statistics are still poor, and it is premature to exclude the
possibility of small $z_*$.  With more data, IceCube may see events at
around $E\simeq 6.3\ {\rm PeV}$ in particular in the case with small
$z_*$.

Here, let us comment on the possibilities to acquire information about
the lifetime of $X$ in the present scenario.  Because the detailed
shape of the spectrum depends on the lifetime, it may be possible to
distinguish the cases with $\tau_X\ll t_0$ and $\tau_X\gg t_0$.  As we
can see, the spectrum smoothly continues to $E>E_{\rm peak}$ if
$\tau_X\ll t_0$.  On the contrary, for $\tau_X\gg t_0$, the neutrino
flux is dominated by the one originating in the Galaxy.  Then, the
spectrum is sharply peaked at $E=\bar{E}_\nu$.  Thus, if the spectrum
of the neutrinos is precisely determined in the future, it will
provide important information about the lifetime of $X$.  Another
possibility is to use the directional information about the high
energy neutrinos.  In the case of $\tau_X\ll t_0$, high energy
neutrinos originate in the decay of $X$ at high redshift so that they
are isotropic.  On the contrary, for the case of $\tau_X\gg t_0$, a
large fraction of the high energy neutrino events are Galactic
origin. Consequently, the neutrino flux is enhanced for the direction
of the Galactic center.  We define $\theta$ as the angle between the
direction of the high-energy neutrino and that of the Galactic center,
with $\theta=0$ being the direction of the Galactic center.  Then, we
calculate $\bar{\Phi}_{\nu, {\rm PeV}}^{\rm
  (Galaxy)}(\theta<90^\circ)$ and $\bar{\Phi}_{\nu, {\rm PeV}}^{\rm
  (Galaxy)}(\theta>90^\circ)$, which are angular-averaged neutrino
fluxes in the regions of $\theta<90^\circ$ and $\theta>90^\circ$,
respectively, for $1<E<1.6\ {\rm PeV}$ (see Eq.\ \eqref{PhiBar}).  For
$\tau_X\gg t_0$, $\bar{\Phi}_{\nu, {\rm PeV}}^{\rm
  (Galaxy)}(\theta<90^\circ)/\bar{\Phi}_{\nu, {\rm PeV}}^{\rm
  (Galaxy)}(\theta>90^\circ)\simeq 2$.  Thus, significant angular
dependence is expected for $\tau_X\gg t_0$, while the flux is
isotropic for $\tau_X\ll t_0$.  Experimental determination of the
directional distribution is important to distinguish the scenarios
with $\tau_X\ll t_0$ and $\tau_X\gg t_0$.

Before closing this letter, several comments are in order.  First, we
consider a possible scenario to produce $X$ in the early universe.  As
we have seen, for $\bar{\Phi}_{\nu, {\rm PeV}} = 3\times 10^{-16} \
{\rm m^{-2}\ sec^{-1}\ str^{-1}\ GeV^{-1}}$, very small value of $Y_X$
is required.  In particular, if $\tau_X\ll t_0$, $Y_X\sim
O(10^{-(26-27)})$ is necessary, which is much smaller than the typical
thermal relic abundance.  Even if we assume that $X$ is in the thermal
bath, however, $Y_X$ can be significantly suppressed with large
entropy production.  A mini inflation after the first inflation (which
is responsible for the density perturbation of the universe) may be an
example.  Another possibility is to introduce a field which has a very
small branching ratio into $X$; if such a field is produced in the
early universe, $X$ can be produced by its decay.

  We also comment on possible constraints from high energy cosmic
  rays, in particular, $\gamma$-ray.  With the mass of $X$ as large as
  $\sim 10^{10}\ {\rm GeV}$, electroweak gauge bosons and charged
  leptons are also produced by electroweak jet cascade even if $X$
  dominantly decays into neutrinos \cite{Berezinsky:2002hq}.  (If $X$
  has decay modes into gauge bosons or into charged leptons, they also
  contribute.)  If too much $\gamma$-ray is generated, the present
  scenario conflicts with the observations of extragalactic
  $\gamma$-ray.  In the present scenario, however, we expect that the
  flux is small enough by estimating the total amout of energy
  injection due to the decay of $X$.  Assuming that the productions of
  electroweak gauge bosons and charged leptons are subdominant
  compared to the neutrino production, the energy density of radiation
  (i.e., $\gamma$-ray) from $X$ should be smaller than
  $E_\gamma^2\Phi_\gamma\lesssim O(10^{-4}\ {\rm m^{-2}\ sec^{-1}\
    str^{-1}\ GeV})$ (see Eq.\ \eqref{PhibarIceCube}).  Notice that,
  if the production process of $\gamma$-ray is suppressed, the flux
  becomes smaller; this is the case for the electroweak jet cascade
  processes.  For the energy range for which the measurement of the
  extragalactic $\gamma$-ray flux is available (i.e.,
  $E_\gamma\lesssim 100\ {\rm GeV}$), the $\gamma$-ray flux in the
  present scenario is found to be smaller than the observed one (which
  is $E_\gamma^2\Phi_\gamma\gtrsim O(10^{-3}\ {\rm m^{-2}\ sec^{-1}\
    str^{-1}\ GeV})$) \cite{Abdo:2010nz}.  Thus, we believe that the
  present scenario is not excluded by the current measurements of high
  energy extragalactic $\gamma$-ray flux.  With future improvements of
  the measurements, signals of the decay of $X$ may be seen.  The
  detailed understanding of the signals requires a precise calculation
  of the spectrum of $\gamma$-ray, which is beyond the scope of this
  letter.

In summary, in this letter, we have studied the cosmic-ray neutrinos
produced by a long-lived particle $X$.  The PeV neutrino events
observed by IceCube may be due to the neutrinos produced by a heavy
particle $X$.  We have discussed that such a scenario works even with
the mass of $X$ much higher than $\sim {\rm PeV}$ if the lifetime is
shorter than the present cosmic time. 
To make such a scenario viable, we have seen that
the decay of $X$ should occur at $z\lesssim O(10^3)$, and that 
the scale of the new physics responsible for the IceCube events
can be as large as $O(10^{9-10}\ {\rm GeV})$.  Detailed study
of the propagation of neutrinos taking into account the effects of the
secondary neutrinos is necessary to understand the case with
$m_X\gtrsim O(10^{9-10}\ {\rm GeV})$.

Finally, we discuss particle-physics models which contain a long-lived
particle decaying into neutrino.  Even assuming that $X$ is a neutral
scalar particle, one may consider the case where $X$ is embedded into
$SU(2)_L$ triplet (with the hypercharge of $+1$)
\cite{Feldstein:2013kka}.  Another possibility is to introduce
$SU(2)_L$ doublet boson (with the hypercharge of $-1/2$) other than
ordinary Higgs boson, which couples to lepton doublet and right-handed
neutrino $\nu_R^c$.  Then, identifying the neutral component of the
doublet as $X$, the decay process $X\rightarrow \nu_L\nu_R^c$ becomes
possible if the neutrino mass is Dirac type or the Majorana mass of
$\nu_R^c$ is small enough.  So far, we have considered the case where
$X$ is neutral.  For the case of $\tau_X\ll t_0$, however, $X$ may be
charged (or even colored) because the constraints on stable superheavy
charged particle (in particular, those using sea water
\cite{Verkerk:1991jf}) do not apply.  From particle-physics point of
view, there exist various beyond-standard-model physics which contain
the candidate of $X$, in particular when the scale of the new physics
is around $O(10^{9-10}\ {\rm GeV})$.  One example is a fermion in
Peccei-Quinn (PQ) sector \cite{Peccei:1977hh, Peccei:1977ur} (or its
supersymmetric partner) in hadronic axion model \cite{Kim:1979if,
  Shifman:1979if}.  The PQ (s)fermion may be stable if it has no
mixing with standard-model fermions.  With introducing a very small
mixing, the PQ (s)fermion decays into standard-model particles with a
very long lifetime.  Similar argument holds for a messenger
(s)fermion in gauge-mediation supersymmetry breaking model
\cite{Dine:1993yw, Dine:1994vc, Dine:1995ag}.  Some of those particles
may play the role of $X$ if they decay into neutrinos at $z\lesssim
O(10^3)$.

\vspace{1em}
\noindent {\it Acknowledgements}: The authors thank S. Matsumoto and
K. Nakayama for useful discussion.  
We are also grateful to A. Kusenko for bringing our attention to the 
astrophysical constraints from high energy cosmic rays. 
The work of Y.E. and R.J. is
supported by the Program for Leading Graduate Schools, MEXT, Japan.
The work of R.J. is also supported by JSPS Research Fellowships for
Young Scientists.  The work of T.M. is supported by Grant-in-Aid for
Scientific research from the Ministry of Education, Science, Sports,
and Culture (MEXT), Japan, No.\ 22244021, No.\ 22540263, and No.\
23104008.

\appendix

\section*{Appendix}

In this Appendix, we summarize the cross sections for the neutrino
scattering processes which are relevant for the calculation of the
damping rate of high energy neutrinos.  The fermion masses (in
particular, lepton masses) are neglected except in Eq.\
\eqref{nunubr2ffbr}.
\begin{itemize}
\item $\nu_\ell + \bar{\nu}_{\ell, {\rm BG}} \rightarrow \nu_\ell +
  \bar{\nu}_\ell$: 
\begin{align}
    \sigma = &
    \frac{g_Z^4}{192\pi} 
    \frac{s}{(s-m_Z^2)^2 + m_Z^2 \Gamma_Z^2}
    + \frac{g_z^4}{64\pi s}
    \left[ x_Z^{-1} + 2 - 2(1+x_Z) \log( 1 + x_Z^{-1} ) \right]
    \nonumber \\ & +
    \frac{g_Z^4}{64\pi} 
    \frac{s}{(s-m_Z^2)^2 + m_Z^2 \Gamma_Z^2}
    (1-x_Z) 
    \left[
      3 + 2x_Z - 2(1+x_Z)^2 \log (1 + x_Z^{-1})
    \right],
  \end{align}
  where, here and hereafter, $x_V\equiv m_V^2/s$ (with $V=W$ and $Z$).
\item $\nu_\ell + \bar{\nu}_{\ell, {\rm BG}} \rightarrow \ell +
  \bar{\ell}$:
  \begin{align}
    \sigma = &
    \frac{g_Z^2 (g_{Z,\ell_L}^2 + g_{Z,\ell_R}^2)}{48\pi} 
    \frac{s}{(s-m_Z^2)^2 + m_Z^2 \Gamma_Z^2}
    + \frac{g_2^4}{16\pi s}
    \left[ x_W^{-1} + 2 - 2(1+x_W) \log( 1 + x_W^{-1} ) \right]
    \nonumber \\ & +
    \frac{g_Z g_{Z,\ell_L} g_2^2}{16\pi} 
    \frac{s}{(s-m_Z^2)^2 + m_Z^2 \Gamma_Z^2}
    (1-x_Z) 
    \left[
      3 + 2x_W - 2(1+x_W)^2 \log (1 + x_W^{-1})
    \right].
  \end{align}
\item $\nu_\ell + \bar{\nu}_{\ell, {\rm BG}} \rightarrow f + \bar{f}$
  ($f\neq\nu_\ell$, $\ell$):
  \begin{align}
    \sigma = &
    \frac{g_Z^2 (g_{Z,f_L}^2 + g_{Z,f_R}^2)}{48\pi}
    \frac{s + 2 m_f^2}{(s-m_Z^2)^2 + m_Z^2 \Gamma_Z^2}
    \sqrt{1 - \frac{4m_f^2}{s}},
    \label{nunubr2ffbr}
  \end{align}
  where $m_f$ is the mass of $f$.
\item $\nu_\ell + \bar{\nu}_{\ell',{\rm BG}}
  \rightarrow \nu_\ell + \bar{\nu}_{\ell'}$ ($\ell\neq\ell'$):
  \begin{align}
    \sigma =
    \frac{g_2^4}{64\pi s}
    \left[ x_Z^{-1} + 2 - 2(1+x_Z) \log( 1 + x_Z^{-1} ) \right].
  \end{align}
\item $\nu_\ell + \bar{\nu}_{\ell',{\rm BG}}
  \rightarrow \ell + \bar{\ell}'$ ($\ell\neq\ell'$):
  \begin{align}
    \sigma =
    \frac{g_2^4}{16\pi s}
    \left[ x_W^{-1} + 2 - 2(1+x_W) \log( 1 + x_W^{-1} ) \right].
  \end{align}
\item $\nu_\ell + \nu_{\ell, {\rm BG}} \rightarrow \nu_\ell + \nu_{\ell}$:
  \begin{align}
    \sigma = &
    \frac{g_Z^4}{64\pi s} \frac{1}{x_Z (1+x_Z)}
    + \frac{g_Z^4}{32\pi s} \frac{1}{1 + 2x_Z}
    \log (1 + x_Z^{-1} ).
  \end{align}
\item $\nu_\ell + \nu_{\ell',{\rm BG}}
  \rightarrow \nu_\ell + \nu_{\ell'}$ ($\ell\neq\ell'$):
  \begin{align}
    \sigma = &
    \frac{g_Z^4}{64\pi s} \frac{1}{x_Z (1+x_Z)}.
  \end{align}
\end{itemize}
In the above expressions, $g_2$ is the gauge coupling constant for
$SU(2)_L$, $g_Z\equiv\sqrt{g_2^2+g_1^2}$ (with $g_1$ being the gouge
coupling constant for $U(1)_Y$), and
\begin{align} 
  g_{Z,u_L} & = \frac{1}{2} g_Z
  - \frac{2}{3} \frac{g_1^2}{g_Z},
  \\
  g_{Z,u_R} &= - \frac{2}{3} \frac{g_1^2}{g_Z},
  \\
  g_{Z,d_L} & = - \frac{1}{2} g_Z
  + \frac{1}{3} \frac{g_1^2}{g_Z},
  \\
  g_{Z,d_R} &= \frac{1}{3} \frac{g_1^2}{g_Z},
  \\
  g_{Z,\ell_L} & = - \frac{1}{2} g_Z
  + \frac{g_1^2}{g_Z},
  \\
  g_{Z,\ell_R} &= \frac{g_1^2}{g_Z},
  \\
  g_{Z,\nu_L} & = \frac{1}{2} g_Z,
  \\
  g_{Z,\nu_R} &= 0.
\end{align}

\end{document}